\documentclass[a4paper]{jpconf}
\usepackage{graphicx}
\usepackage{physics}
\usepackage{breqn}
\usepackage{amsmath}
\newcommand{\PT}{\mathcal{PT}}
\newcommand{\be}{\begin{equation}}
\newcommand{\ee}{\end{equation}}

\begin{document}
\title{$\PT$ Symmetry and Renormalisation in Quantum Field Theory} 

\author{Carl M Bender$^1$, Alexander Felski$^2$, S P Klevansky$^3$,
and Sarben Sarkar$^4$}

\address{$^1$Department of Physics, Washington University, St Louis, Missouri
63130, USA }
\ead{cmb@wustl.edu}
\vspace{.4cm}
\address{$^2$Institut f\"ur Theoretische Physik, Universit\"at Heidelberg,
69120 Heidelberg, Germany} 
\ead{felski@thphys.uni-heidelberg.de}
\vspace{.4cm}

\address{$^3$Institut f\"ur Theoretische Physik, Universit\"at Heidelberg, 
69120 Heidelberg, Germany}
\ead{spk@physik.uni-heidelberg.de}
\vspace{.4cm}
\address{$^4$Department of Physics, King's College London, London WC2R 2LS, UK}
\ead{sarben.sarkar@kcl.ac.uk}

\begin{abstract}

Quantum systems governed by non-Hermitian Hamiltonians with $\PT$ symmetry are
special in having real energy eigenvalues bounded below and unitary time
evolution. We argue that $\PT$ symmetry may also be important and present at the
level of Hermitian quantum field theories because of the process of
renormalisation. In some quantum field theories renormalisation leads to
$\PT$-symmetric effective Lagrangians. We show how $\PT$ symmetry may allow
interpretations that evade ghosts and instabilities present in an interpretation
of the theory within a Hermitian framework. From the study of examples
$\PT$-symmetric interpretation is naturally built into a path integral
formulation of quantum field theory; there is no requirement to calculate
explicitly the $\PT$ norm that occurs in Hamiltonian quantum theory. We discuss
examples where $\PT$-symmetric field theories emerge from Hermitian field
theories due to effects of renormalization. We also consider the effects of
renormalization on field theories that are non-Hermitian but $\PT$-symmetric
from the start.
\end{abstract}

\section{Introduction}
Non-Hermitian Hamiltonians govern systems that in general receive energy from
and/or dissipate energy into their environment and so they are typically not in
equilibrium. Their energy is not conserved and their energy levels are complex.
However, in 1998 \cite{R1a} non-Hermitian $\PT$ systems were shown to offer new
possibilities for unitary time evolution in quantum mechanics. Since then there
has been extensive research activity \cite{R2a}, particularly in material science and optics,
which has implemented the ideas of $\PT$-symmetric quantum mechanics. 

$\PT$-symmetric quantum mechanics can be considered to be a one-dimensional
quantum field theory. To date there have been no clear examples of
$\PT$-symmetric quantum field theory in higher dimensions. We examine the
possibility that $\PT$ symmetry may emerge when considering the effect of
quantum fluctuations in a higher-dimensional field theory. We also examine the
converse process where we consider the effect of quantum fluctuations on an
initially $\PT$-symmetric quantum field theory.

Using numerical techniques the work of Bender and Boettcher \cite{R1a} showed
that quantum-mechanical Hamiltonians of the form
\begin{equation}
\label{e1}
H=\frac{1}{2}p^{2}+x^{2}\left(ix\right)^{\epsilon}
\end{equation}
have a real positive spectrum. Using a correspondence between ordinary
differential equations and integrable field theory models, Dorey {\it et al.}
\cite{R3a} provided an analytical proof of the result of Bender and Boettcher. 

Mostafazadeh \cite{R4a} considered the framework of pseudo-Hermiticity which
contained $\PT$ symmetry as a special case. A Hamiltonian $H$ is
pseudo-Hermitian if $H$ is not selfadjoint but
\begin{equation}
\label{e2}
H^\dagger=\eta H\eta^{-1},
\end{equation}
where $\eta$ is a positive-definite Hermitian operator. In terms of $\eta$ the
Hilbert space of states has a positive-definite inner product given by
 \begin{equation}
\label{e3}
\left\langle \varphi,\eta\chi\right\rangle
\end{equation}
where $\left\langle,\right\rangle$ is a conventional inner product on the
Hilbert space of states for the Hermitian part of the Hamiltonian. There is not
a universal expression for $\eta$, which is difficult to calculate in general.
This would seem to imply an impediment to calculating correlation functions in
$\PT$-symmetric field theories. Within the context of quantum mechanics,
formulated in terms of functional integrals, Jones and Rivers \cite{R5a} showed
through examples that it was not necessary to calculate $\eta$. Hence, we shall
take the path-integral representation as the definition of a quantum field
theory and, except for the Lee model, not consider $\eta$ explicitly. It is a
convenient formulation for both perturbative and nonperturbative calculations.

We consider four examples of field theories to illustrate three different
aspects of $\PT$ symmetry:
\begin{enumerate}
\item the Lee model and the elimination of ghosts \cite{R6a};
\item the role of the top quark and the Higgs instability \cite{R7a} in the
standard model of particle physics \cite{R8a};
\item the new epsilon expansion for $\PT$-symmetric scalar field theory
\cite{R9a,R10a};
\item the functional renormalisation group and the preservation of $\PT$
symmetry in the infrared after renormalisation \cite{R11a}.
\end{enumerate} 

\section{Role of $\PT$ symmetry in the Lee model}
Hermitian quantum field theory (QFT) has been studied for many decades both
perturbatively and nonperturbatively. QFT is considered to be a viable and
successful description of quantum electrodynamics, the weak, and the strong
interactions. The use of dispersion relations in QFT, provides a way to bypass
some of the limitations of perturbation theory and was developed decades ago
(starting with the work of Lehmann, Symanzik, and Zimmerman \cite{R12a, R12aa}).
The theory of dispersion relations is based on one particularly important
assumption: There exists an interpolating (i.e. a renormalised) field $\phi(x)$,
where $\phi(x)$ denotes an arbitrary field of the theory, such that we have the weak convergence relation
\begin{equation}
\label{e4 }
\lim_{t\to\pm\infty}\mel{\alpha}{\phi (x)-\phi_{\rm out,\,in}(x)}{\beta}=0,
\end{equation}
where $\ket{\alpha}$ and $\ket{\beta}$ are experimentally accessible states and
$\phi_{\rm out}$ and $\phi_{\rm in}$ are free field operators associated with
noninteracting particles. The Green's functions in dispersion relations are
assumed to have been made finite by some form of renormalisation (which need not
be specified). In this framework it is possible, within the context of any
specified polynomial field-theory Lagrangian, to establish coupled integral
equations among 1-particle irreducible Green's functions. For example, in a
Yukawa theory of a pseudoscalar field interacting with fermions, 2-point
functions are coupled with the 3-point (vertex) function. One conclusion is that
such a theory may contain ghosts unless the vertex function vanishes as the
momentum transfer tends to infinity \cite{R12aa}. So it is quite possible that a
Hermitian quantum field theory may contain ghosts. The Lee model \cite{R13a},
which we discuss next, is just such an example where the vertex function and in
particular coupling-constant renormalisation is related to such unusual
behaviour. The role of $\PT$ symmetry is crucial in banishing the ghosts and
making the field theory viable \cite{R6a}. 

\subsection{Lee model}
The Lee model is very different from the Standard Model of particle physics but
it has the advantage of being soluble and so provides a testing ground for new
methods in field theory. Indeed, the model was devised as a field theory whose
coupling-constant and wave-function renormalisation could be computed exactly in
principle. One variant of the model contains fields for infinitely heavy
spinless fermions, which have two internal states $V$ and $N$, as well as a
neutral scalar field $\theta$. Antifermions are not in the theory and so there
is no crossing symmetry. The permitted reaction channel is
\begin{equation}
\label{e5 }
V\rightleftharpoons N+\theta,
\end{equation}
but the channel
\begin{equation}
\label{e6 }
N\rightleftharpoons V+\theta
\end{equation}
is not allowed.

In momentum space the Hamiltonian of the Lee model (in a space of finite volume
$\mathcal{V}$) is
\begin{equation}
\label{e7 }
H=H_{0}+H_{\rm int},
\end{equation}
where 
\begin{equation}
\label{e8}
H_{0}=m_{V}\overline{\psi}_{V}{\psi}_{V}+m_{N}\overline{\psi}_{N}{\psi}_{N}+\sum_{\bf k} \omega_{\bf k}a^{\dag}_{\bf k}a_{\bf k}
\end{equation}
and 
\begin{equation}
\label{e9}
H_{\rm int}=\delta m_{V}\overline{\psi}_{V}{\psi}_{V}-g_{0}{\mathcal{V}}^{-\frac{1}{2}}
\sum_{\bf k} \frac{u(\omega_{\bf k})}{{{(2\omega_{\bf k})}}^{\frac{1}{2}}}
(\overline{\psi}_{V}{\psi}_{N}a_{\bf k}+\overline{\psi}_{N}{\psi}_{V}
a^{\dag}_{\bf k}).
\end{equation}
$u(\omega_{\bf k})$ is a dimensional cut-off function  which is chosen to tend to $0$ for large $\omega_{\bf k}$ where $\omega_{\bf k}=\sqrt{{\bf k}^{2}+\mu^{2}}$.
Standard commutation and anti-commutation rules are assumed:
\begin{align}
\label{ }
\{\overline{\psi}_{V},{\psi}_{V}\}& =\{\overline{\psi}_{N},{\psi}_{N}\}=1\\
\left[a_{\bf k}, a^{\dag}_{\bf k'}\right]& = \delta_{\bf k, \bf k'}.
\end{align}

All other commutators and anticommutators vanish. Bare operators and coupling
constant appear in $H$. However $m_V,\,m_N,\,{\rm and},\,\mu$ are renormalised
parameters and so are determined by experiment; $\delta m_V$ is the mass
counterterm for the $V$ particle and is a function of $g_0$. No further mass
renormalisation is actually needed. The coupling $g_0$ in principle is
determined from the scattering cross-section of $N$ and $\theta$ although there
are interesting complications of non-Hermiticity if the coupling is not small.

However, let us for the moment persist in the view that $g_0$ is real and so $H$
is Hermitian. Let us choose as a basis for the Hilbert space states of the form
\begin{equation}
\label{ }
\ket{}=\ket{n_{V},n_{N},\{{n_{\bf {k}}}\}}.
\end{equation}
From $H$ it is clear that there are two conserved quantities $B$ and $Q$:
\begin{equation}
\label{ }
B=n_V+n_N, \quad  Q=-N_\theta+n_N,
\end{equation}
where $N_\theta=\sum_{\bf k} n_{\bf k}$. Hence the Hilbert space is partitioned
into an infinite number of independent sectors with fixed $B$ and $Q$. Although
this makes the model soluble, the analysis of sectors with large $B$ and $Q$ is
complicated. Since our main point is to show how renormalisation can lead to a
non-Hermitian Hamiltonian, we simplify our analysis further by considering (i) a
nontrivial sector $B=1$ and $Q=0$ and (ii) modifying the model so that there is
no $\bf k$ dependence. This simplified model is a quantum-mechanical one since
the quantum fields in (\ref{e8}) and (\ref{e9}) are replaced by quantum
operators. Hence, infinities that arise from summing over an infinite set of
modes in quantum field theory are absent but some features of coupling-constant
renormalisation persist. The resulting Hamiltonian is $\mathcal{H}={\cal{H}}_{0}
+{\cal{H}}_{1}$, where 
\begin{equation}
\label{e14}
{\cal{H}}_{0}={m_{V}}V^{\dag} V+m_{N}N^{\dag}N+\mu a^{\dag} a
\end{equation}
and 
\begin{equation}
\label{e15}
{\cal{H}}_{1}=g_{0}(V^{\dag}Na+a^{\dag}N^{\dag}V) +\delta m_{V} V^{\dag} V.
\end{equation}
We will look at the 2-dimensional Hilbert space associated with this sector and
consider the eigenstates, which we will denote by 
\begin{align}
\label{e16}
\ket{V}& =c_{11}\ket{1,0,0}+c_{12}\ket{0,1,1},\\
\ket{N\theta}& =c_{21}\ket{1,0,0}+c_{22}\ket{0,1,1},
\end{align} 
with eigenvalues $m_V$ and $E_{N\theta}$. Let us denote $(m_{V}+\delta m_{V})$
by $m_{V_0}$, the bare mass of $V$. The eigenvalues can be shown to satisfy the equations
\begin{align}
\label{e17}
m_{V}=& \frac{1}{2} (m_{N}+m_{\theta}+m_{V_0} -\sqrt{M_{0}^{2}+4g_{0}^{2}}),\\
E_{N\theta}=& \frac{1}{2}(m_{N}+m_{\theta}+m_{V_0}+\sqrt{M_{0}^{2}+4g_{0}^{2}}),
\end{align}
where $M_{0}\equiv m_{N}+m_{\theta}-m_{V_0}.$ Following field theory, we define
the wave-function renormalisation constant $Z_{V}$ through the relation
\begin{equation}
\label{e18}
1=\bra{0}\frac{1}{\sqrt{Z_{V}}}V\ket{V},
\end{equation}
which gives
\begin{equation}
\label{e19}
Z_{V}=\frac{2g_{0}^{2}}{\sqrt{M_{0}^{2}+4g_{0}^{2}}\,(\sqrt{M_{0}^{2}+4g_{0}^{2}
}-M_{0})},
\end{equation}
and the coupling constant renormalisation through
\begin{equation}
\label{e20}
\frac{g^{2}}{g_{0}^{2}}=Z_{V}.
\end{equation}
We deduce that 
\begin{equation}
\label{e21}
g_{0}^{2}=\frac{g^{2}}{1-\frac{g^{2}}{M^{2}}},
\end{equation}
where $M \equiv  m_{N} +m_{\theta}-m_{V}$. If $g$, the experimentally determined
value, exceeds $M$ then $g_{0}$ becomes pure imaginary and the Hamiltonian
becomes non-Hermitian. Although the Hamiltonian has become non-Hermitian, it is
$\PT$-symmetric. Explicitly, the transformations due to $\mathcal P$ are
\cite{R6a}
\begin{align}
\label{e22}
\mathcal PV\mathcal P& =-V, & \mathcal PN\mathcal P& =-N, &
\mathcal Pa\mathcal P& =-a, \\  
\mathcal PV^{\dag}\mathcal P& =-V^{\dag}, & 
\mathcal PN^{\dag}\mathcal P& =-N^{\dag}, & 
\mathcal Pa^{\dag}\mathcal P& =-a^{\dag}.
\end{align}
The transformations due to $\mathcal T$ are
\begin{align}
\label{e23}
\mathcal TV\mathcal T& =V, & \mathcal TN\mathcal T& =N, & 
\mathcal Ta\mathcal T&= a\\
\mathcal TV^{\dag}\mathcal T& =V^{\dag}  &  
\mathcal TN^{\dag}\mathcal T& =N^{\dag} &
\mathcal Ta^{\dag}\mathcal T& = a^{\dag}.   
\end{align}
It is now straightforward to check that $i|g_{0}|(V^{\dag}Na+a^{\dag}N^{\dag}V)$
is $\PT$-symmetric. This is our first example of a $\PT$-symmetric field theory
that arises from a Hermitian field theory after renormalisation. On introducing
a $\PT$-symmetric inner product in the ghost regime of the Lee model, the
so-called ghost state (identified within the Hermitian frame work) turns out to
have a positive norm\cite{R6a}. The Lee model then can be interpreted as an acceptable
quantum field theory.

\section{Higgs instability and $\PT$ symmetry}
The conventional approach to renormalisation involves the regularisation of loop
integrals in Feynman diagrams. This led to scale dependence of the parameters of
the theory which can be understood in a different way using the approach \cite{R15a} of
Wilson to renormalisation. Wilson's approach leads naturally to the concept of
effective actions, and the form of the effective actions for some theories in
high-energy physics is non-Hermitian and $\PT$-symmetric. 

The change from the conventional to the Wilson approach can be illustrated by
considering a scalar field $\Phi$ whose (conventional) action is given by
\begin{equation}
\label{e24}
S_{\Lambda}\left[\Phi\right]=\int d^{D}x (\frac{1}{2}\partial_{\mu}\Phi\partial^{\mu}\Phi+U_{\Lambda}(\Phi))
\end{equation}
where the UV cut-off is $\Lambda$. If we integrate over Fourier modes with
momenta of magnitude $p$ in the shell $k\le p \le \Lambda$ we arrive at an
action which we will denote by $S_{k}\left[\phi\right]$ where $\phi$ has Fourier
modes with $p \le k$ (a sharp infrared cut-off).\footnote{Wetterich~\cite{R16a} introduced
a smooth infrared cut-off function $R_k(p^2)$ rather than a sharp cut-off so
that $S_k\left[\phi\right]=\int d^{D}x [\frac{1}{2}\partial_\mu\phi\partial^\mu+
V_k(\phi)]+\Delta_{k}[\phi]$, where $\Delta_{k}[\phi]=\frac{1}{2}\int d^{D}p
\phi_{p}R_{k}(p^{2})\phi_{-p}$. A common choice for $R_k(p^2)$ is $R_k(p^{2})=
(k^{2}-p^{2})\Theta(k^2-p^2)$.} The associated partition function is
\begin{equation}
\label{e26}
Z_k[j]\equiv \exp(-W_{k}[j]) \equiv \int D\phi \exp(-S_{k}[\phi]-\int_{p}j_{p}\phi_{-p})
\end{equation}
and a Legendre transformation on $W_k[j]$ leads to the effective action $\Gamma_{k}[\phi_{c}]$
\begin{equation}
\label{e27}
\Gamma_{k}[\phi_{c}]=W_{k}\left[j\right]-\int d^{D}x j\phi_{c}-\Delta_{k}[\phi_{c}].
\end{equation}
Within systematic (derivative) approximation schemes \cite{R16a, R17a}  the effective action can be
represented (at the lowest level of approximation) by the {\it ansatz}
\begin{equation}
\label{e28}
\Gamma_{k}[\phi_{c}]=\int d^{D}x (\frac{1}{2}\partial_{\mu}\phi_{c}\partial^{\mu}\phi_{c}+U_{k}(\phi_{c})).
\end{equation}
The effective potential, $U_{k}(\phi_{c})$ which results from renormalisation, will show the emergence of $\PT$ symmetry in the next  models that we will discuss. This ansatz represents quite a severe approximation since, for an $x$-independent $\phi_{c}$, we have a one-dimensional approximation to an infinite-dimensional field.

We consider two theories of current interest using this formalism:
\begin{enumerate}
\item a theory of dynamical breaking of gravity via  a graviton condensate field \cite{R18a}
$\varphi_{c}$ with $U(\varphi_{c})=-{\varphi_{c}}^{4}\log(i \varphi_{c})$ for
large $\varphi_{c}$r,  
\vspace{.2cm}
  
and 
  
\item the Standard Model of particle physics for which $\varphi_{c}$ is the
Higgs field and $U(\varphi_{c})=-{\varphi_{c}}^{4}\log(\varphi_{c}^{2})$ for
large $\varphi_{c}$ \cite{R7a}.
\end{enumerate}
\vspace{.2cm} 
 
Both examples, in conventional quantum mechanics, would show unstable behaviour
for large $\varphi_{c}$. Under $\mathcal P$ we have $\varphi_{c}\rightarrow
{-\varphi_{c}}$ and under $\mathcal T$ we have $i\rightarrow{-i}$. So $U(
\varphi_{c})$ is $\PT$ symmetric in both cases.
 
We study such effective potentials by considering three related quantum
mechanical Hamiltonians $H_{i}$, $i=1,2,3$:
 \begin{align}
\label{e29}
   H_{1} = & p^{2}+{x^{4}} \log (ix),  \\
    H_{2} =&  p^{2}-{x^{4}} \log (ix), \\
    H_{3}=& p^{2}-x^{4}\log(x^{2}),
\end{align}
which are non-Hermitian but $\PT$-symmetric. We will find that the the first two Hamiltonians will show unbroken $\PT$ symmetry and the Hamiltonian $H_{3}$ will show broken $\PT$ symmetry.

\subsection{The spectra for $H_{1}$ }
We use the WKB method of semiclassical quantum mechanics to determine the energy
spectrum of $H_1$:
\begin{itemize}
\item Locate turning points.
\item Examine the complex classical trajectories on an infinite-sheeted
Riemann surface.
\item Determine the open or closed nature of the trajectories.
\end{itemize}
The turning points satisfy the equation
\be
\label{E6}
E=x^{4}\log (ix).
\ee
We take $E=1.24909$ because this is the numerical value of the ground-state
energy obtained separately by solving the Schr\"odinger equation. (See the table
below.)

One turning point lies on the negative imaginary-$x$ axis. To find this point we
set $x=-ir$ ($r>0$) and obtain the algebraic equation $E=r^{4}\log r$. Solving
this equation by using Newton's method, we find that the turning point lies at
$x=-1.39316i$. To find the other turning points we seek solutions to (\ref{E6})
in polar form $x=re^{i\theta}$ ($r>0,\,\theta\,{\rm real}$). Substituting for
$x$ in (\ref{E6}) and taking the imaginary part, we obtain 
\begin{equation}
\log r=-(2k\pi+\theta+\pi/2)\cos(4\theta)/\sin(4\theta),
\label{E7}
\end{equation}
where $k$ is the sheet number in the Riemann surface of the logarithm.
(We choose the branch cut to lie on the positive-imaginary axis.)
Using (\ref{E7}), we simplify the real part of (\ref{E6}) to 
\begin{equation}
E=-r^{4}(2k\pi+\theta+\pi/2)/\sin(4\theta).\label{E8}
\end{equation}

We then use (\ref{E7}) to eliminate $r$ from (\ref{E8}) and use Newton's method
to determine $\theta$. For $k=0$ and $E=1.24909$, two $\PT$-symmetric
(left-right symmetric) pairs of turning points lie at $\pm0.93803-0.38530i$ and
at $\pm0.32807+0.75353i$. For $k=1$ and $E=1.24909$ there is a turning point at
$-0.53838+0.23100i$; the $\PT$-symmetric image of this turning point lies on
sheet $k=-1$ at $0.53838+0.23100i$.

\begin{figure}[h]
\begin{center}
\includegraphics[width=2in]{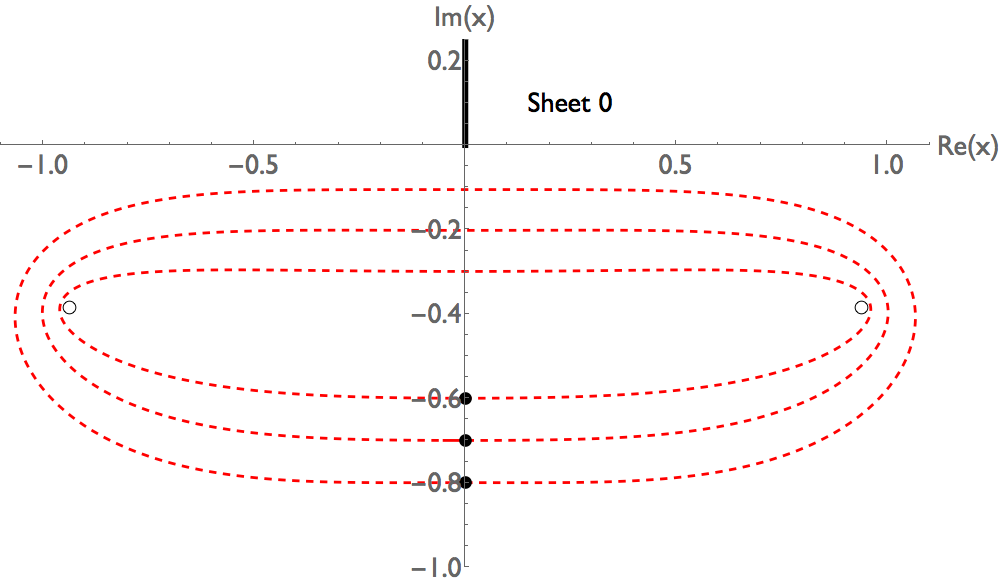}
\caption{Three nested closed classical paths.}
\label{F1}
\end{center}
\end{figure}

\begin{figure}[h]
\begin{center}
\includegraphics[width=2in]{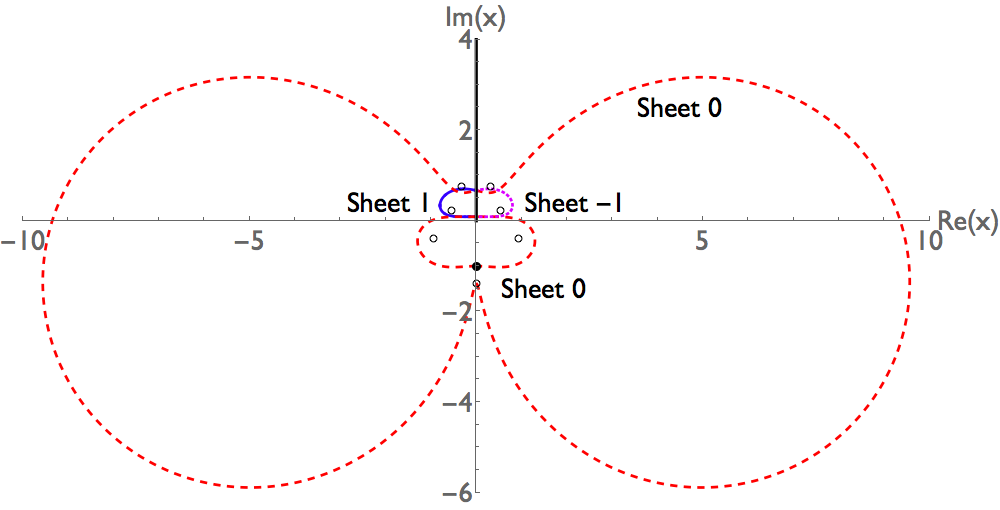}
\caption{{\it Closed} classical path for energy $E=1.24909$.}
\label{F2}
\end{center}
\end{figure}

The turning points determine the shape of the classical trajectories. Two
topologically different kinds of classical paths are shown in Figs.~\ref{F1} and
\ref{F2}. All classical trajectories are {\it closed} and {\it left-right
symmetric}, and this implies that the quantum energies are all real \cite{R19a}.

The WKB quantization condition is a complex path integral on the principal
sheet of the logarithm ($k=0$). On this sheet a branch cut runs from
the origin to $+i\infty$ on the imaginary axis; this choice of branch
cut respects the $\PT$ symmetry of the configuration. The integration
path goes from the left turning point $x_{{\rm L}}$ to the right
turning point $x_{{\rm R}}$ \cite{R20a}: 
\begin{equation}
\left(n+\frac{1}{2}\right)\pi\sim\int_{x_{{\rm L}}}^{x_{{\rm R}}}dx\sqrt{E-V(x)}\quad(n>>1).\label{E9}
\end{equation}

If the energy is large $\left(E_{n}\gg1\right)$, then from (\ref{E7}) with $k=0$
we find that the turning points lie slightly below the real axis at $x_{\rm R}=
re^{i\theta}$ and at $x_{{\rm L}}=re^{-\pi i-i\theta}$
with 
\begin{equation}
\theta\sim-\pi/(8\log r)\quad{\rm and}\quad r^{4}\log r\sim E.\label{E10}
\end{equation}
We choose the path of integration in (\ref{E9}) to have a constant
imaginary part so that the path is a horizontal line from $x_{{\rm L}}$
to $x_{{\rm R}}$. Since $E$ is large, $r$ is large and thus $\theta$
is small. We obtain the simplified approximate quantization condition
\begin{equation}
\left(n+\frac{1}{2}\right)\pi\sim r^{3}\log r\int_{-1}^{1}dt\sqrt{1-t^{4}},\label{E11}
\end{equation}
which leads to the WKB approximation for $n\gg1$: 
\begin{equation}
\frac{E_{n}}{[\log(E_{n})]^{1/3}}\sim\left[\frac{\Gamma(7/4)(n+1/2)\sqrt{\pi}}{\Gamma(5/4)\sqrt{2}}\right]^{4/3}.\label{E12}
\end{equation}

\begin{tabular}{ |p{3cm}|p{3cm}|p{3cm}|p{3cm}|p{3cm}| }
\hline
\multicolumn{5}{|c|}{Calculation of values of energy} \\
\hline
Energy level $n$ & $E_n$& $\frac{E_n}{[\log(E_n)]^{1/3}}$ & WKB & \% error \\
\hline
0  & 1.24909 & 2.06161 & 0.54627 & 73.5028    \\
3&   13.7383& 9.96525 & 7.31480 & 26.5969  \\
6 &31.6658 & 20.9458 & 16.6979 & 20.2804\\
9    &52.9939 & 33.4674 & 27.6956 & 17.2463\\
12 &   76.9748& 47.1776 & 39.9324 & 15.3573 \\
\hline
\end{tabular}

\begin{figure}[h]
\begin{center}
\includegraphics[width=2in]{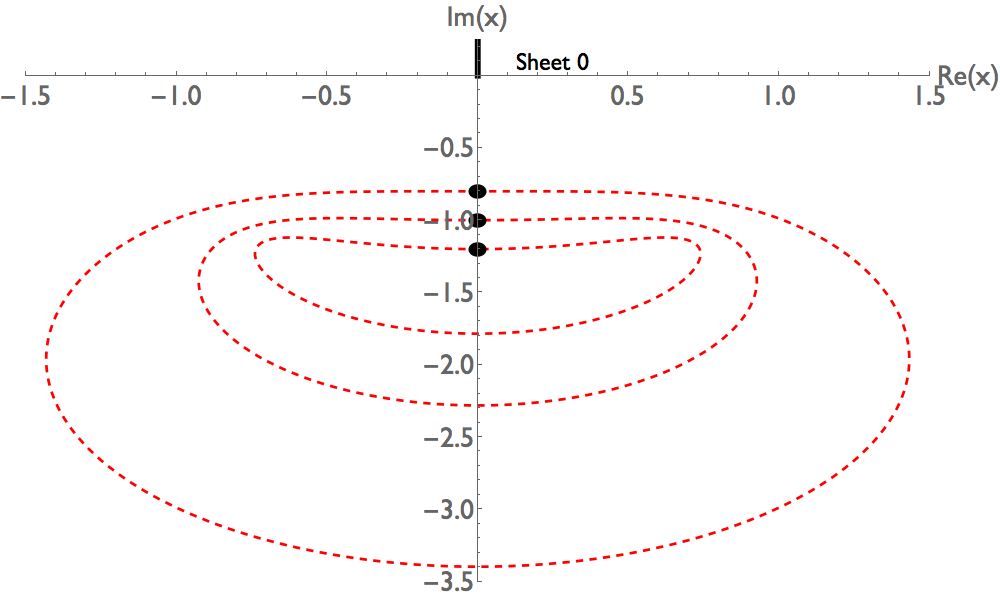}
\caption{Three nested classical trajectories for $H_2$ with $E=2.07734$.}
\label{F3}
\end{center}
\end{figure}

\begin{figure}[h]
\begin{center}
\includegraphics[width=2in]{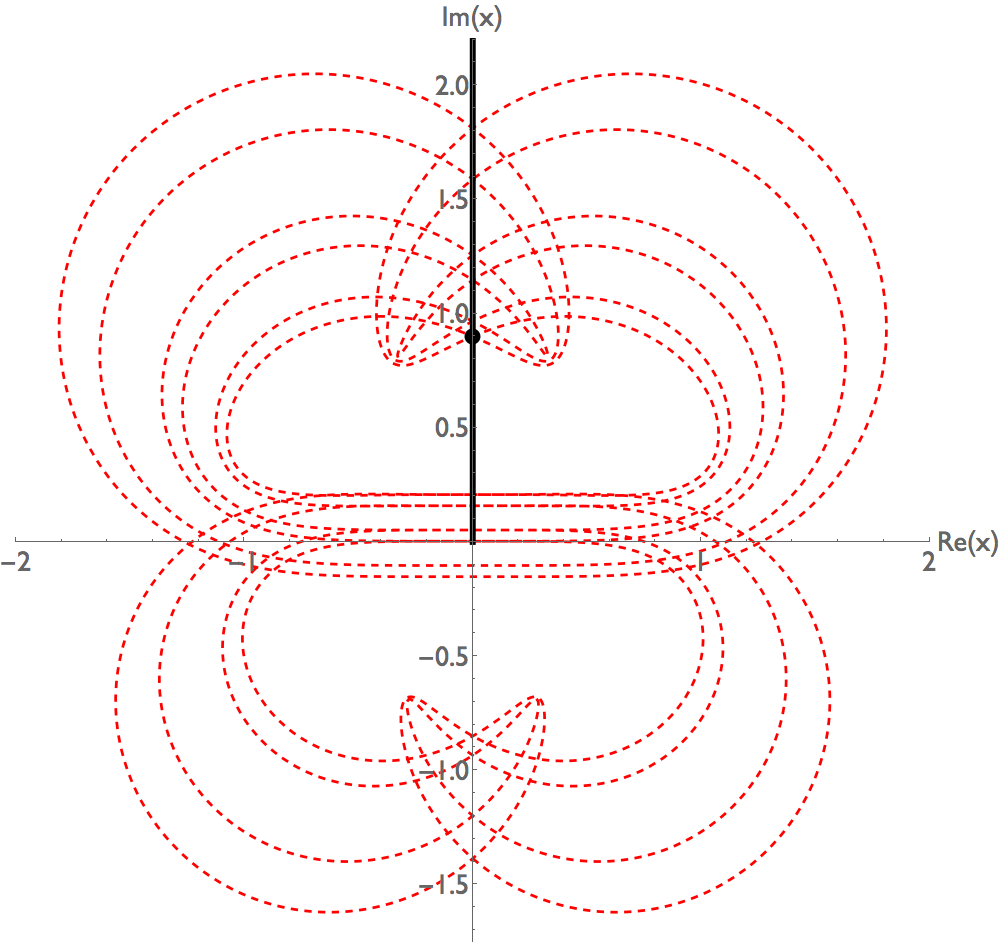}
\caption{Complex classical trajectory for $H_2$ with $E=2.07734$.}
\label{F4}
\end{center}
\end{figure}

\noindent{{\bf Analysis of the supergravity model Hamiltonian} $H_2$:}
The classical trajectories for the Hamiltonian $H_2$ are plotted
in Figs.~\ref{F3} and \ref{F4}. Like the classical trajectories
for the Hamiltonian $H_{1}$), these trajectories are closed, which
implies that all the eigenvalues for $H_{2}$  are real.

\vspace{.2cm}

\begin{figure}[h]
\begin{center}
\includegraphics[width=2in]{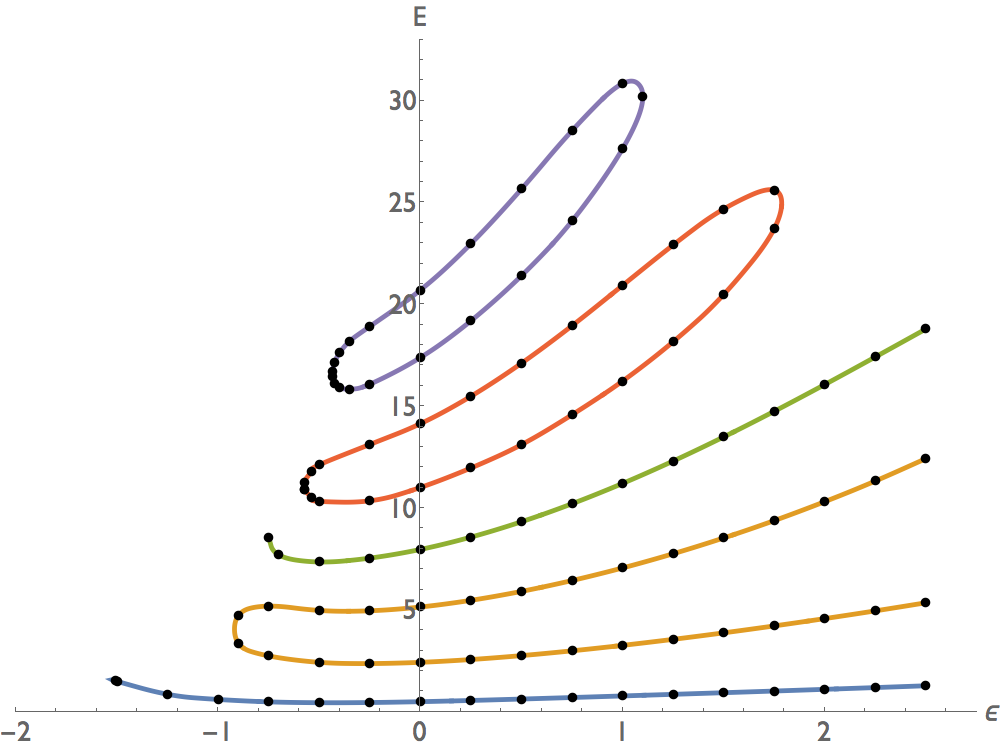}
\caption{{\it Energies of the Hamiltonian} $H=p^{2}+x^{2}(ix)^{\epsilon}
\log(ix)$ plotted versus $\epsilon$.}
\label{F6}
\end{center}
\end{figure}

\begin{figure}[h]
\begin{center}
\includegraphics[width=2in]{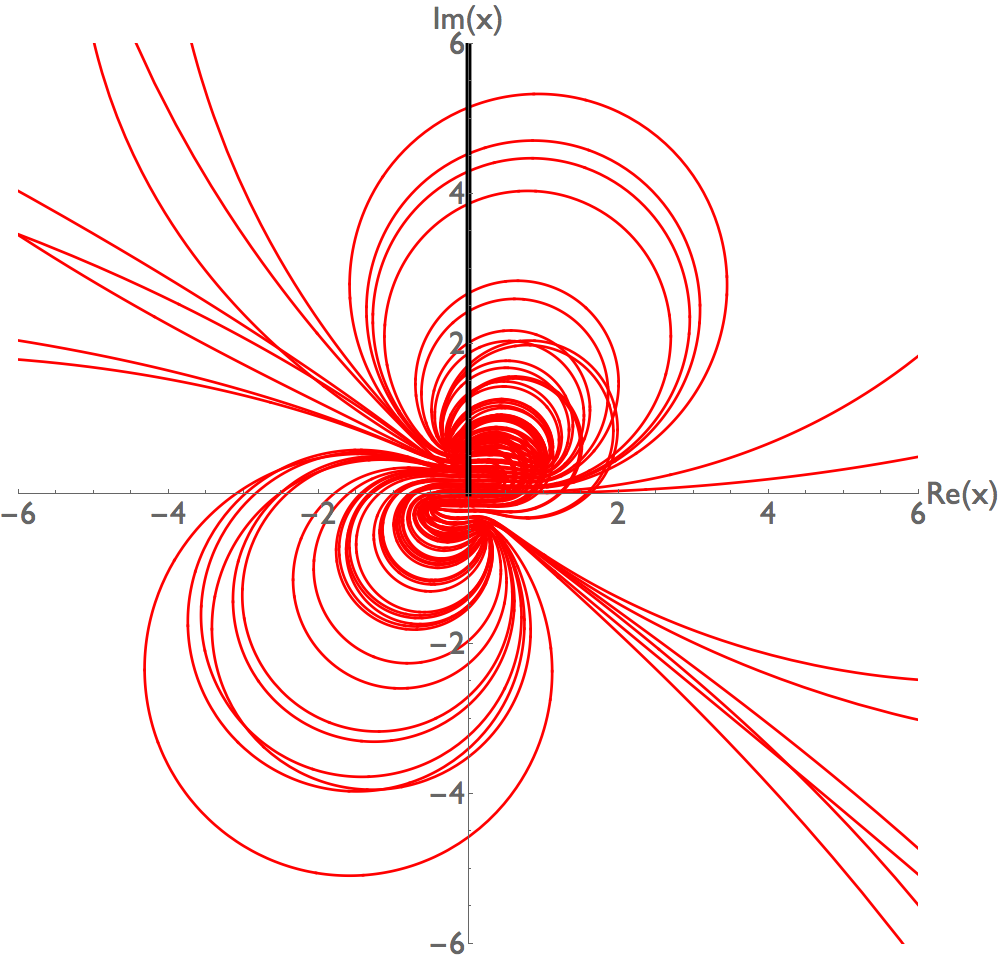}
\caption{{\it Classical path for the Hamiltonian} $H=p^{2}-x^{4}\log\left(x^{2}
\right)$.}
\label{F7}
\end{center}
\end{figure}

\noindent {\bf Analysis of the Higgs model Hamiltonian $H_3$:}
To make sense of $H_3$ we again introduce a parameter $\epsilon$ and we define
$H_3$  as the limit of $H=p^{2}+x^{2}(ix)^{\epsilon}\log\left(x^{2}\right)$ as
$\epsilon:\,0\to2$. This case is distinctly different from that for $H_2$.
Figure~\ref{F6} shows that the $\PT$ symmetry is broken for all
$\epsilon\neq0$. When $\epsilon=2$, there are only four real eigenvalues:
$E_{0}=1.1054311$, $E_{1}=4.577736$, $E_{2}=10.318036$, and $E_{3}=16.06707$.
To confirm this result we plot a classical trajectory for $\epsilon=2$
in Fig.~\ref{F7}. In contrast with Fig.~\ref{F4}, the trajectory
is open and not left-right symmetric.

\section{Renormalisation and canonical scalar field theory} 
We have shown examples of $\PT$-symmetric quantum field theories arising from
the process of renormalisation in field theories. The properties of
$\PT$-symmetric field theories have yet to be established. In quantum mechanics
much progress in understanding $\PT$ symmetry has been made through the study of
the Hamiltonian in (\ref{e1}). A natural generalisation to an Euclidean field
theory Hamiltonian in $D$-dimensions is to consider the Lagrangian
\be
\label{e30}
L=\frac{1}{2}(\nabla \phi )^{2}+\frac{1}{2}\mu^{2}\varphi^{2}+\frac{1}{2}g\mu_{0}^{2}\phi^{2}(i\mu_{0}^{\frac{\delta}{2}}\phi)^{\epsilon}
\ee
where $\phi$ is a dimensional pseudoscalar field, $\delta=2-D$ and $\epsilon\ge
0$. It is natural to study this class of field theories since we can then
compare our findings in certain limits with those found in quantum mechanical
systems. For noninteger $\epsilon$ the interaction is nonpolynomial and the
methods that are normally used for calculating Greens functions in quantum field
theory do not apply. The method we will use to set up a quantum field theory
involves
\begin{enumerate}
\item rewriting of the interaction in terms of a formal series of polynomial
interactions;
\item creating modified Feynman rules in a nonperturbative expansion;
\item consideration of infinite contributions and renormalisation.
\end{enumerate}
The first part of the method, used in (i), was introduced decades ago and has
been recently developed in (ii) and (iii) within the context of $\PT$-symmetric
field theory. The procedure of (i) allows us to use Wick's theorem and the
linked-cluster theorem in conjunction with (ii).

In a conventional field theory described by a Lagrangian $\mathcal L$ within
polynomial interactions in a field $\varphi$, we calculate connected Green's
functions using a partition functional $Z(J)$ defined through a path integral as
\be
\label{e31}
Z(J)=\int {\mathcal{D}} \varphi \exp(-\int d^{D}x ({\mathcal{L}}+J\varphi)),
\ee
where $J$ is a source. It is well known that the normalised partition functional
$\frac{Z(J)}{Z(0)}$ satisfies
\be\ln
\label{e32}
\frac{Z(J)}{Z(0)}=\exp(\sum{\rm Connected\,source-to-source\,diagrams}).
\ee
This result simplifies our calculations in (ii) and (iii). First, we note that
formally, on writing $\psi=\mu_{0}^{\frac{\delta}{2}}\varphi$ (and suppressing
the argument of $\psi$)
\be
\label{e33}
\psi^{2}(i\psi)^{\epsilon}=\sum_{n=0}^{\infty}\frac{\epsilon^{n}}{n!}\sum_{r=0}^{n}\left(\begin{array}{c}n \\r\end{array}\right)\Big(\frac{i\pi \vert \psi \vert}{\psi}\Big)^{r}\psi^{2}\Big( \frac{1}{2} \ln \psi^{2}\Big)^{n-r}=\sum_{n=1}^{\infty}\epsilon^{n}\mathcal{L}_{n}+ \psi^{2}
\ee
because
\be
\label{e34}
\ln(i\psi)=\frac{1}{2}i\pi \vert \psi \vert /\psi  +\frac{1}{2} \ln \psi^{2}.
\ee
The series in (\ref{e33}) can be expressed in terms of powers of $\psi$ on
noting that 
\begin{enumerate}
\item 
\label{e50}
\be
\Big(\frac{ \vert \psi \vert}{\psi}\Big)^{r}=\sum_{w=0}^{\infty}
\mathcal{I}_{w}\psi^{(2w+1)r},
\ee
where $\mathcal{I}_{w}=\frac{2}{\pi}\int_0^\infty dt\frac{(-t^2)^{w}}{(2w+1)!}$.
\item 
\be
\label{e51}
(\ln \psi^{2})^{s}=\mathcal{F}^{0}_{N,s}\psi^{2N},
\ee
where $\mathcal{F}^{0}_{N,s}=\lim_{N \to 0}(\frac{d}{dN})^{s}$.
\end{enumerate}

Thus, we can rewrite the partition function in (\ref{e31}) as
\be
\label{e35}
Z[J]=\int {\mathcal{D}} \varphi \exp(-\int d^Dx (\sum_{n=0}^\infty\epsilon^{n}
\mathcal{L}_{n}+J\varphi)),
\ee
where $\mathcal{L}_0$ is the free (that is, $\epsilon$-independent) part of the
Lagrangian. All $n$-point Green's functions can be evaluated as
$$\frac{\delta^{n}Z[J]}{\delta J(x_{1})\delta J(x_{2})\cdots \delta J(x_{n})}\Big\vert_{J=0}=\int {\mathcal{D}} \varphi \exp(-\int d^{D}x \sum_{n=0}^{\infty}\epsilon^{n}\mathcal{L}_{n} )
\varphi (x_{1}) \cdots \varphi (x_n).$$
We are only interested in the  connected Green's functions $G_{c}(x_{1},x_{2},
\cdots,x_{n})$. Expanding the integrand in the path integral in powers of
$\epsilon$ we obtain products of $\mathcal{L}_{n}$ which consist of integrals of
powers of $\phi$ and the operations denoted by $\mathcal{I}_{w}$ and
$\mathcal{F}^0_{N,s}$. On passing the operators $\mathcal{I}_w$ and $\mathcal{
F}^0_{N,s}$ from inside the path integral to outside the path integral, we are
left with a path integral which can be evaluated using Wick's theorem. This is a
well-defined procedure (explored in two recent papers \cite{R9a, R10a}) which has the advantage
that the path integral is performed along the real $\phi$-axis. Thus, we need not be
concerned with the hopelessly  complicated (infinite-dimensional) integration
paths that terminate in complex Stokes sectors. 

\subsection{Results to ${\rm O}(\epsilon)$}
In general $Z=\exp(-E_{0}V)$ where $E_{0}$ is the ground-state energy density
and $V$ is the spacetime volume. Using the above method to $O(\epsilon)$ we find
for general $D$ that
\be
\label{e53}
\Delta E=\frac{1}{4}\epsilon (4\pi)^{-D/2}\Gamma(1-\frac{1}{2}D) \Big\{ \ln[2(4\pi)^{-D/2}\Gamma(1-\frac{1}{2}D)]+\psi(3/2) \Big\}.
\ee
As a check, for $D=0$ and first order in $\epsilon$, we have
\begin{equation}
\Delta E=\frac{\epsilon}{2\sqrt{2\pi}}\int_{-\infty}^\infty d\phi\exp(-\frac{1}
{2}\phi^{2})\phi^{2}\ln(i \phi)=\frac{\epsilon}{4}[\psi(3/2)+\ln(2)],
\end{equation}
which agrees with (\ref{e53}). For $D=1$, $\Delta E$ is the expectation of the
interaction Hamiltonian to $O(\epsilon)$ in the unperturbed (Gaussian) ground
state
\begin{equation}
\label{}
\Delta E=\frac{\epsilon}{2}\int_{-\infty}^\infty dx\exp(-x^{2}) x^{2}\ln(ix)
\Big/\int_{-\infty}^{\infty} dx\exp(-x^{2})=\frac{\epsilon}{8} \psi(3/2).
\end{equation}
    
The calculated higher order connected Green's function to $O(\epsilon)$ is
\be
\label{e57}
G_c(y_1,\cdots,y_n)=-\frac{1}{2}\epsilon(-i)^n\Gamma(\frac{n}{2}-1)[\frac{1}{2}
(4\pi)^{-D/2}\Gamma(1-\frac{1}{2}D)]^{1-n/2}\int d^Dx\prod_{k=1}^n\triangle_{1}
(y_{k}-x),
\ee
where the free propagator $\triangle_{\lambda}(x)$ is associated with $L_{0}=
\frac{1}{2}(\nabla \phi)^{2}+\frac{1}{2}\lambda^{2}\phi^{2}$ and obeys the
equation
\be
\label{e58}
(-\nabla^{2}+\lambda^{2})\triangle_{\lambda}(x)=\delta^{(D)}(x).
\ee
The solution of $(\ref{e58})$ is 
\be
\label{e59}
\triangle_{\lambda}(x)=\lambda^{D/2-1}\vert x \vert^{1-D/2}(2\pi)^{-D/2}K_{1-D/2}(\lambda \vert x \vert)
\ee
and
\be
\label{}
\triangle_{\lambda}(0)=\lambda^{D-2} (4\pi)^{-D/2} \Gamma(1-\frac{D}{2})\sim \frac{1}{2\pi\delta}  \,\quad (\delta \to 0).
\ee
From (\ref{e57}) it is clear that as $\delta\to 0+$, the connected Green's
functions $G_{c}(y_{1},\cdots,y_{n})\to 0$ for $n \ge 3$. This indicates that at
least to ${\rm O}(\epsilon)$ the theory is noninteracting at $D=2$. When we
consider ${\rm O}(\epsilon^2)$ contributions, we will reexamine this issue.
  
Turning to $n=1$ and $n=2$ we have
\be
\label{61}
G_{1}=-i\epsilon \sqrt{\frac{1}{2}\pi(4\pi)^{-D/2} \Gamma(1-\frac{D}{2})}
\ee
and the two-point function ${\widetilde{G}}_{2}(p)$ in momentum space is
\be
  \label{62}
  {\widetilde{G}}_{2}(p)=1/[p^{2}+1+\epsilon K+O(\epsilon^{2})]
  \ee
  where $K=\frac{3}{2}-\frac{1}{2}\gamma +\frac{1}{2}\ln[\frac{1}{2}(4\pi)^{-D/2}\Gamma(1-\frac{D}{2})]$. Thus the renormalised mass to $O(\epsilon)$ is 
\be
\label{}  
M_{R}^{2}=1+K\epsilon+O(\epsilon^{2}).
\ee
So near $D=2$
\be
\label{}
G_{1}\sim -i\epsilon \frac{1}{2\sqrt \delta}
\ee
and 
\be
\label{}
M_{R}^{2}\sim-\frac{1}{2}\epsilon\ln\delta +A,
\ee
where $A=1+\epsilon[\frac{3}{2}-\frac{\gamma}{2}-\frac{1}{2}\ln(4\pi)]$. Because of the singularities in $G_{1}$ and $M_{R}^{2}$ as $\delta \to 0$ some renormalisation is needed to remove these infinities. The question is whether perturbative renormalisation can be performed in the context of the novel $\epsilon$-expansion method. We can remove the divergence in $G_1$ by introducing in the Lagrangian a linear counter term $iv\phi$ where $v$ has dimension $(\rm{mass})^{1+D/2}$ and $v=v_{1} \epsilon +v_{2} \epsilon^{2}+v_{3}\epsilon^{3}+\cdots$. Since $v$ is real such a term is compatible with $\PT$-symmetry. Adding also a mass counter term $\mu$ we can consider the Lagrangian density
\be
\label{}
L=\frac{1}{2}(\nabla \phi)^{2}+\frac{1}{2}\mu^{2}\phi^{2}+\frac{1}{2}g
\mu_{0}^{2}\phi^{2}(i\mu_{0}^{1-D/2}\phi)^{\epsilon},
\ee
where $\phi$ is dimensionful. Using this Lagrangian we obtain
\be
\label{}
G_{1}=-\frac{i\epsilon g}{m^2}\mu_{0}^{D/2-1}\sqrt{\frac{1}{2}\pi m^{D-2}\triangle_{1}(0)}+\frac{i\epsilon v_{1}}{\mu_{0}^{2}m^{2}}
\ee
and for the renormalised mass
\be
\label{}
M_{R}^{2}=(m\mu_{0})^{2}+\frac{1}{2}\epsilon g\mu_{0}^{2}\{3-\gamma+\ln [\frac{1}{2} m^{D-2}\triangle_{1}(0)]\}.
\ee
In both expressions we have introduced the dimensionless quantity $m^{2}=g+\mu^{2}/\mu_{0}^{2}$ and as $\delta \to 0$
\be
\label{}
G_{1}\sim \frac{i\epsilon}{g\mu_{0}^{2}+\mu^{2}}(v_{1}-\frac{g\mu_{0}^{2}}{2\sqrt{\delta}})
\ee
and
\be
\label{}
M_{R}^{2}\sim\mu^{2}-\frac{1}{2}\epsilon g \mu_{0}^{2}\ln \delta +A,
\ee
where $A=g\mu_{0}^{2}\{1+\epsilon [\frac{3}{2}-\frac{\gamma}{2}-\frac{1}{2}
\ln(4 \pi)]\}$ is a finite quantity. By setting $v_{1}=\frac{g\mu_{0}^2}{2
\sqrt{\delta}}$ we have a finite $G_{1}=0$. $M_{R}$ is logarithmically divergent
in $\delta$. We absorb this divergence into $\mu$ by setting
\be
\label{}
\mu^{2}=B+\frac{\epsilon}{2}g\mu_{0}^{2}\ln\delta,
\ee
so that $M_R^2=A+B$ and $B$ is a finite quantity determined in principle
from experiment.

\subsection{Calculations to second order in $\epsilon$}
In second order the calculations become much more involved. The ${\rm O}(
\epsilon^2)$ contribution to the connected part of $G_1$, which we denote by
$G_{1,2}$, can be shown to be
\begin{multline}
\label{}
G_{1,2}=-\frac{1}{2}igm^{-2}\sqrt{\frac{1}{2}\pi \triangle(0)}
\Big\{[\ln[2\mu_{0}^{2-D}\triangle(0)]+\psi(2)\Big\}+
\frac{1}{8}ig^{2}m^{-4}\sqrt{\frac{1}{2}\pi \triangle(0)}
\Big\{(\ln[2\mu_{0}^{2-D}\triangle(0)]\\
+\psi(\frac{3}{2}))(6-D)+2D-4+ 4\triangle(0)\mu_{0}^{2}m^{2}\int d^{D}x (1+
\frac{\triangle(x)}{\triangle(0)})^{2} \ln[1+\frac{\triangle(x)}{\triangle(0)}]
\Big\},
\end{multline}
where $\triangle(0)$ denotes for brevity $\triangle_{\mu_{0}m}(0)$. As $\delta
\to 0$,
\begin{equation}
G_{1,2}\sim-\frac{i}{4}gm^{-2}\delta^{-1/2}[\psi(2)-\ln(\pi \delta)]
+\frac{i}{4}g^{2}m^{-4}\delta^{-1/2}[1+\psi(3/2)-\ln \pi-\ln \delta]+O(\delta).
\end{equation}
So the algebraic divergence $\delta^{-1/2}$ persists to $O(\epsilon^2)$. The
divergence can be removed through $v_2$. Similarly the $\ln\delta$ divergence
persists for $G_2$ at second order. Interestingly, the higher-order Green's
functions continue to vanish for $D=2$. Avoidance of a noninteracting theory
remains an open question in this approach~\cite{R10a}.

\section{Renormalisation group flows of $\PT$-symmetric theories}
So far we have considered whether a Hermitian theory can lead to a non-Hermitian
$\PT$-symmetric theory due to the effect of renormalisation. We ask the opposite
question in this section: Can a $\PT$-symmetric field theory retain its $\PT$
symmetry as the Lagrangian flows due to renormalisation?

We review some preliminary work on this question. In its full generality this is
an intractable problem. We turn to the framework of the {\it
functional renormalization group} \cite{R15a}, which combines the functional
formulation of quantum field theory with the Wilsonian renormalization group. It
is possible to make some progress in solving the functional equations using the simpler approach developed by
Wetterich~\cite{R16a} and Morris~\cite{R17a} which, in the local potential approximation
($\ref{e28}$), leads to a nonlinear {\it partial differential equation} (PDE)
rather than a functional equation. This simplification enables substantial
progress in understanding the effective potential in arbitrary spacetime
dimensions. In $D$ dimensions this PDE is
\begin{equation}
\partial_kU_k\left(\phi_c\right)=\frac{1}{\pi_D}\frac{k^{D+1}}{k^2+U_k''\left(
\phi_c\right)},
\label{Er8}
\end{equation}
where $\pi_D=\frac{D(2\pi)^D}{S_{D-1}}$ and $S_{D-1}=\frac{2\pi^{D/2}}{\Gamma
(D/2)}$ is the surface area of a unit $D$-dimensional sphere. 
We may assume that the equations for $U_k\left(\phi_c\right)$ involve
dimensionless quantities. If this were not so, we could achieve
dimensionlessness by introducing a mass scale $M$. For example, for $D=1$ the
dimensionless variables, denoted by a tilde, are $\tilde{\phi}=M^{1/2}\phi$,
$\tilde{g}=M^{-3}g$, $\tilde{k}=M^{-1}k$, and $\tilde{\mu}=M^{-1}\mu$. The
Wetterich equation (\ref{Er8}) can be thought of as being in terms of such
dimensionless variables.

To avoid difficulties in numerical analysis associated with boundary conditions,
in the past the PDE (\ref{Er8}) has been analyzed by approximating
$U_k\left(\phi_c\right)$ as a finite series in powers of the field $\phi_c$.
This {\it ansatz} leads to a sequence of coupled nonlinear {\it ordinary}
differential equations. The consistency of such a procedure has not been established.

For $D=1$ (the quantum-mechanical case), (\ref{Er8}) becomes
\begin{equation}
\partial_kU_k\left(\phi_c\right)=\frac{1}{32\pi^2}\frac{k^2}{k^2+U_k''\left(
\phi_c\right)}.
\label{Er9}
\end{equation}
Even in this one-dimensional setting, no exact solution to this nonlinear PDE is
known and only numerical solutions have been discussed. 

We depart from the above treatment by performing an asymptotic analysis for
large values of the cut-off $k$~\cite{R11a}. The novelty of the approach used here is that
it avoids the appearance of coupled nonlinear ordinary differential equations.
To leading order the results of this analysis are qualitatively different
depending on whether the space-time dimension $D$ is greater or less than $2$.

Letting $z=k^{2+D}$ and $U_k(\phi_c)=U(z,\phi)$, we can rewrite (\ref{Er8}) as 
\begin{equation}
U_z(z,\phi)=\frac{1}{(2+D)\pi_D}\frac{1}{z^{\frac{2}{2+D}}+U_{\phi\phi}(z,
\phi)},
\label{Er10}
\end{equation}
where the subscripts on $U$ indicate partial derivatives. We now assume that
for large $z$ we can neglect the $U_{\phi\phi}$ term in the denominator. (The
consistency of this assumption is easy to verify when $D<2$.) Then, for large
$z$ we have to leading order in our approximation scheme
\begin{equation}
U_z(z,\phi)\sim\frac{1}{(2+D)\pi_{D}}z^{-\frac{2}{2+D}}\quad(z\gg1).
\label{Er11 }
\end{equation}
On incorporating a correction $\epsilon$ to this leading behavior
\begin{equation}
U(z,\phi)=\frac{1}{D\pi_D}z^{\frac D{2+D}}+\epsilon(z,\phi),
\label{Er12}
\end{equation} 
we get to order $O(\epsilon^2)$
\begin{equation}
\epsilon_z(z,\phi)=-\frac{1}{(D+2)\pi_D}z^{-\frac{4}{2+D}}\epsilon_{\phi\phi}
(z,\phi).
\label{Er13}
\end{equation}
On making the further change of variable $t=\frac{D+2}{2-D}z^{\frac{D-2}{D+2}}$,
(\ref{Er13}) becomes 
\begin{equation}
\epsilon_t(t,\phi)=\frac{1}{(D+2)\pi_D}\epsilon_{\phi\phi}(t,\phi).
\label{Er14}
\end{equation}
The variable $t$ is positive for $D<2$ and negative for $D>2$ and is not defined
at $D=2$. Thus, (\ref{Er14}) is a conventional diffusion equation for $D<2$ but
is a {\it backward} diffusion equation for $D>2$. The backward diffusion
equation is an inverse problem that is ill-posed. The problems
associated with this ill-posedness may
be connected with difficulties in solving (\ref{Er8}) numerically when $D=4$.

In the preliminary study these issues were not addressed further. The case $D
=1$ was considered and some simple $\PT$-symmetric theories were studied~\cite{R11a}. From
(\ref{Er8}), on defining $\widehat{U}_k\equiv U_k-\frac{k^D}{D\pi_D}$,
we can deduce that 
\begin{equation}
\partial_k\widehat{U}_k=-\frac{k^{D-1}\widehat{U}_k''}{\pi_D\left(k^2+
\widehat{U}_k''\right)}.
\label{E18A}
\end{equation}
 From (\ref{E18A}) it is consistent to assume that $\widehat{U}\to V(\phi)$ and
$\frac{\widehat{U}^{''}}{k^2}\to0$ as $k\to\infty$. (For simplicity of
notation we have dropped the suffix $k$ in $\widehat{U}_k$.) Let us write the
correction as $V(\phi)+\frac{1}{k}U_1(\phi)$. On substituting in (\ref{E18A}),
we obtain 
\begin{equation}
U_1(\phi)=\frac{1}{\pi}V''(\phi).
\label{Er19}
\end{equation}
We can proceed in this way and write the next correction as $\widehat{U}(\phi)=
V(\phi)+\frac{1}{k\pi}V''(\phi)+\frac{1}{k^2}U_2(\phi)$. This leads to $U_2(
\phi)=\frac{1}{2\pi^2}V^{(4)}$ where $V^{(4)}(\phi)\equiv\frac{d^4}{d\phi^4}V(
\phi)$. Repeating this procedure, we get $U_3(\phi)=\frac{1}{3}\left(\frac{1}
{2\pi^3}V^{(6)}(\phi)-{V^{(2)}(\phi)}^2\right)$. This procedure can be
formalized: On writing $\delta=\pi/k$ (not to be confused with $\delta$ in the
last section) and $x=\pi\phi$, (\ref{E18A}) becomes 
\begin{equation}
\frac{\partial}{\partial\delta}\widehat{U}=\frac{\frac{\partial^2}{\partial x^2}
\widehat{U}}{1+\delta^2\frac{\partial^2}{\partial x^2}\widehat{U}}
\label{Er20}
\end{equation}
and $\widehat{U}=\sum_{n=0}^{\infty}\delta^nU_n(\phi)$. 

\begin{figure}[h!]
\begin{center}
\includegraphics[scale=0.27]{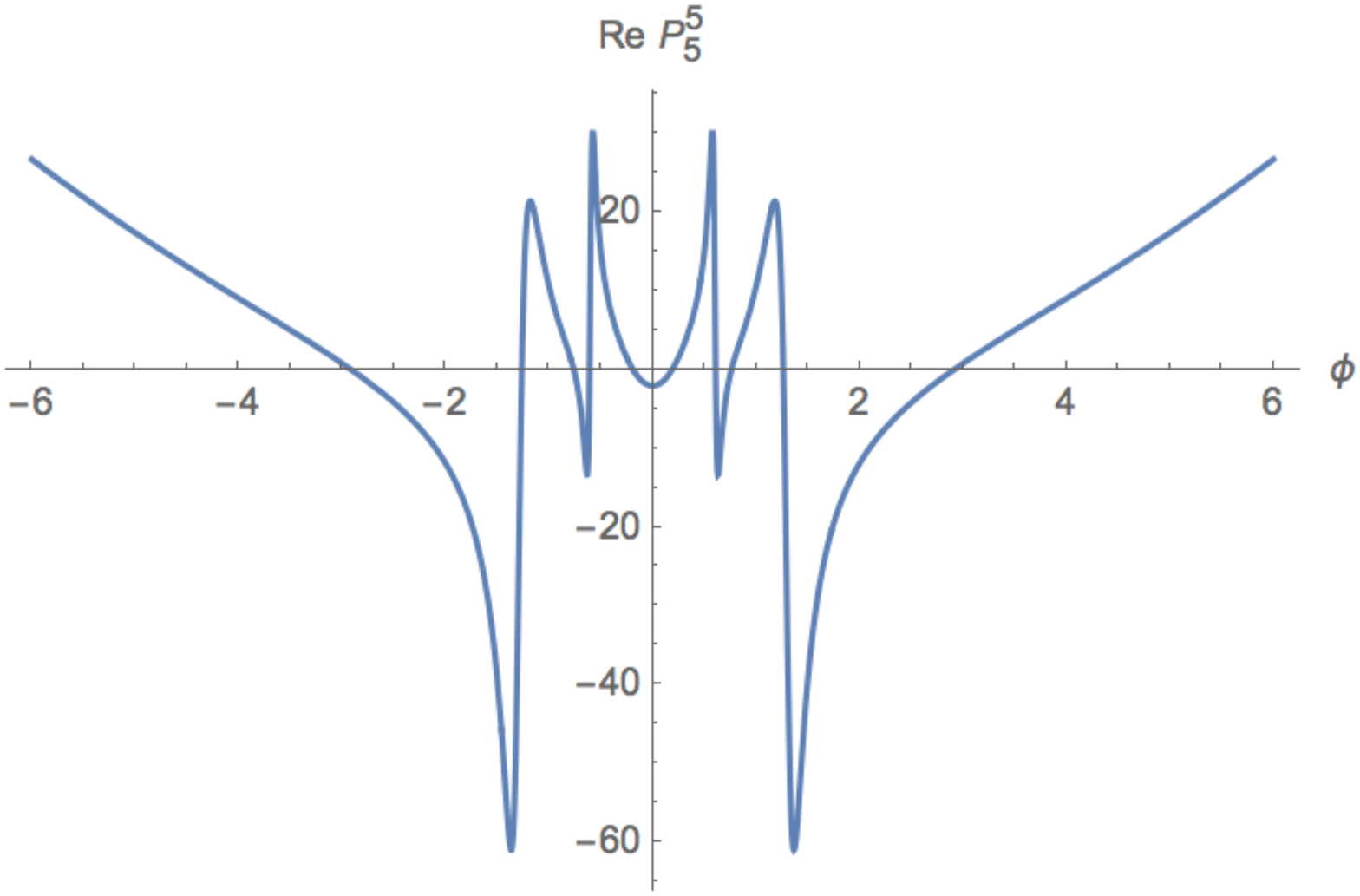}\hspace{.1in}
\includegraphics[scale=0.27]{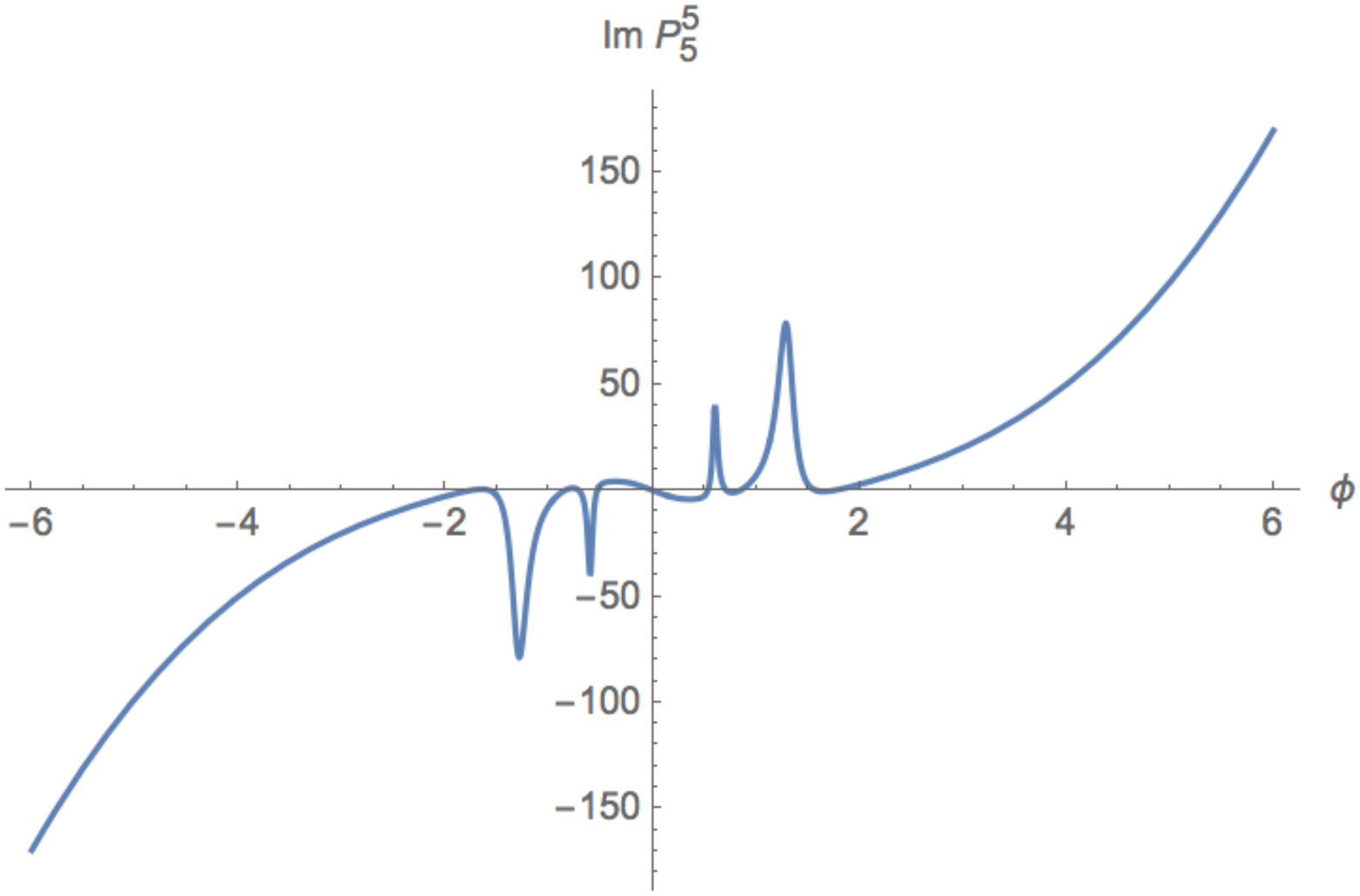}
\end{center}
\caption{Effective potential flow for massive $i\phi^3$ with the mass parameter
$\mu=1$. Shown is a plot of the real and imaginary parts of the $P_5^5$
approximant plotted as functions of real $\phi$. Observe that there are no
poles. The real part of the potential is right-side-up, so there is no
instability. Furthermore, apart from small fluctuations near the origin, the
imaginary part of the potential behaves like $i\phi^3$ for large $|\phi|$.}
\label{f1}
\end{figure}

When $\delta$ is small, the scale $k$ is large and we are probing the
microscopic potential. As $\delta\to\infty$ we probe the infrared limit of the
effective potential. However, the analysis outlined above was based on a
perturbation theory in $\delta$. There is a parallel to the theory of phase
transitions, where a physical entity is expanded in inverse powers of the
temperature $T$, and then an extrapolation procedure is used to find the
critical behavior at lower $T$. This often-used extrapolation technique is
based on Pad\'e approximation.

So far, our treatment has been for a general potential; we will now specialize
$V(x)$ to two cases $V(x)=gx^3$ and $V(x)=gx^4$, which are $\PT$-symmetric for
$g=i$ and $g$ real. We also include a mass term. Our purpose is to consider the
series $\sum_{n=0}^N\delta^nU_n(x)$ generated from the series relevant to these
two cases and to consider the $\delta\to0$ limit.

\subsection{Cubic potentials}
We consider both massless and massive cubic potentials. In the former case
$U_0(\phi)=g\phi^3$ and in the latter case $U_0(\phi)=\mu^2\phi^2+g\phi^3$. 
For the massless case the solution to (\ref{Er20}) is 
\begin{equation}
\begin{array}{ll}
U_1(\phi)=6g\phi,&\qquad\quad U_{10}(\phi)=
\textstyle{\frac{1621728}{175}}g^5\phi^3,\\
U_2(\phi)=0,&\qquad\quad
U_{11}(\phi)=\textstyle{\frac{17882208}{1925}}g^5\phi
-\textstyle{\frac{46656}{11}}g^6\phi^6,\\
U_3(\phi)=-12g^2\phi^2,&\qquad\quad 
U_{12}(\phi)=-\textstyle{\frac{30608064}{385}}g^6\phi^4,\\
U_4(\phi)=-6g^2,&\qquad\quad
U_{13}(\phi)=\textstyle{\frac{279936}{13}}g^7\phi^7
-\textstyle{\frac{4537967328}{25025}}g^6\phi^2,\\
U_5(\phi)=\textstyle{\frac{216}{5}}g^3\phi^3,&\qquad\quad
U_{14}(\phi)=\textstyle{\frac{4512754944}{7007}}g^7\phi^5
-\textstyle{\frac{5900745888}{175175}}g^6,\\
U_6(\phi)=\textstyle{\frac{456}{5}}g^3\phi,&\qquad\quad
U_{15}(\phi)=\textstyle{\frac{1439488599552}{525525}}g^7\phi^3
-\textstyle{\frac{1679616}{15}}g^8\phi^8,\\
U_7(\phi)=-\textstyle{\frac{1296}{7}}g^4\phi^4,&\qquad\quad
U_{16}(\phi)=\textstyle{\frac{295324339824}{175175}}g^7\phi
-\textstyle{\frac{25063743168}{5005}}g^8\phi^6,\\
U_8(\phi)=-\textstyle{\frac{34668}{35}}g^4\phi^2,&\qquad\quad
U_{17}(\phi)=\textstyle{\frac{10077696}{17}}g^9\phi^9
-\textstyle{\frac{105103706989824}{2977975}}g^8\phi^4,\\
U_9(\phi)=\textstyle{\frac{7776}{9}}g^5\phi^5
-\textstyle{\frac{89496}{315}}g^4,&\qquad\quad
U_{18}(\phi)=\textstyle{\frac{3212310887424}{85085}}g^9\phi^7
-\textstyle{\frac{101014620538752}{2127125}}g^8\phi^2.
\end{array}
\label{Er21}
\end{equation}
These expressions, which are valid for pure imaginary $g$ as well as for real
$g$, are cumbersome but manageable. We stress that there has been no truncation
of the function space on which $\widehat{U}(\phi)$ has support. This contrasts
with the usual approach which requires a
truncation at the onset of the calculation of the renormalization group flow.
This absence of truncation continues to be a feature for the massive case.

The iterative solution to (\ref{Er20}) for the massive case is similar to that
for the massless case, but the expressions for $U_{n}(\phi)$ have many more
terms. For example, the coefficient of $\delta^9$ is
\begin{eqnarray}
U_9(\phi)&=&\textstyle{\frac{8}{315}}\big(34020g^5\phi^5+56700g^4\mu^2\phi^4
-11187g^4+37800g^3\mu^4\phi^3\nonumber\\
&&~~+12600g^2\mu^6\phi^2+2100g\mu^8\phi+140\mu^{10}\big),\nonumber
\end{eqnarray}
which in the massless limit $\mu\to0$ reduces to the two-term expression in
(\ref{Er21}). We refrain from listing the coefficients explicitly and instead
proceed directly to the large-$\delta$ behavior of the diagonal Pad\'e
approximants. We denote the diagonal Pad\'e approximants in the massless case by
$P_N^{0,N}(\delta)$ and in the massive case by $P_N^N(\delta)$.
Let us examine some low-order Pad\'e approximants in the
limit $\delta\to\infty$. For example,
$$\lim_{\delta\to\infty}P_2^2(\delta)=\frac{18g^3\phi^5+3g^2\left(10\mu^2\phi^4
-9\right)+14g\mu^4\phi^3+2\mu^6\phi^2}{2\left(3g\phi+\mu^2\right)^2}.$$
[As a check, when $\mu=0$ this expression agrees with the corresponding
expression $g\phi^3-\frac{3}{2\phi^2}$ for $P_2^{0,2}(\delta)$.] For $N=3$ we
obtain
$$\lim_{\delta\to\infty}P_3^{0,3}(\phi)=\frac{g\phi^3\left(736g\phi^5+1705
\right)}{25-800g\phi^5}\qquad{\rm and}\qquad P_3^3(\phi)=-u_3/d_3,$$
where
\begin{eqnarray}
u_3&=&1609632g^8\phi^8+729g^7\phi^3\left(5888\mu^2\phi^4+5115\right)
+243g^6\mu^2\phi^2\left(23008\mu^2\phi^4+15345\right)\nonumber\\
&&~~+1296g^5\mu^4\phi\left(3376\mu^2\phi^4+945\right)+15120g^4\mu^6\left(142
\mu^2\phi^4+9\right)+658944g^3\mu^{10}\phi^3\nonumber\\
&&~~+121824g^2\mu^{12}\phi^2+12288g\mu^{14}\phi+512\mu^{16},\nonumber
\end{eqnarray}
$$d_3=225g^2[7776g^5\phi^5+81g^4(160\mu^2\phi^4-3)+8640g^3\mu^4
\phi^3+2880g^2\mu^6\phi^2+480g\mu^8\phi+32\mu^{10}].$$
[As a check, we have $\lim_{\mu\to0}P_3^3(\phi)=P_3^{0,3}(\phi)$.]

Note that for the massless case $P_2^{0,2}$ has a double pole for both real and
imaginary $g$. In fact, the same is true for $P_n^{0,n}$ for even $n$. This pole
is an artifact of the Pad\'e approximation and is not present when $n$ is odd,
so we will only consider the behavior of these approximants for odd $n$. In
general, the odd-$n$ diagonal Pad\'e approximants have no singularities at all
on the real-$\phi$ axis when $g$ is imaginary  but singularities occur for the case of real $g$ .
These findings indicate that the $\PT$-symmetric effective potential is well
behaved in the infrared limit . From the expressions for the
diagonal Pad\'e approximants we see that for large $|\phi|$, the leading
behavior of the imaginary part of the effective potential is exactly $i\phi^3$.
Consequently the $\PT$ nature of the interaction is preserved under
renormalization.

A similar investigation of quartic potentials also shows that $\PT$-symmetry is maintained under renormalization~\cite{R11a}.

\section{Conclusions}
The study of $\PT$-symmetric quantum field theory is still in its infancy. We
have shown that there is a link between renormalization of Hermitian field
theories and  $\PT$-symmetric field theories. This provides a motivation for the
study of $\PT$-symmetric field theories. Often, the usual tools of quantum field
theory cannot be applied directly. We have reviewed some interesting and
promising approaches to studying these new field theories. Clearly, there remain
some unresolved issues with these approaches, and this will lead to many
opportunities for future research. 

\vspace{.3cm}
\subsection*{Acknowledgments}
\noindent
CMB thanks the Alexander von Humboldt and Simons Foundations for financial
support. SS and CMB thank the UK Engineering and Physical Sciences Research
Council for financial support.

\end{document}